\documentclass{article}

\usepackage{spconf,amsmath,graphicx,amsfonts}
\usepackage{times}
\usepackage{amsbsy}
\usepackage{latexsym}
\usepackage{amssymb}

\usepackage{cite}

\usepackage{bm}
\usepackage{extarrows}

\usepackage{algorithm}
\usepackage{algorithmic}

\newcommand{\reffig}[1]{Fig. \ref{#1}}

\newcommand{\refsec}[1]{Section \ref{#1}}

\makeatletter

\newcommand{\Rmnum}[1]{\expandafter\@slowromancap\romannumeral #1@}
\makeatother

\usepackage{footnote}
\makesavenoteenv{tabular}
\newcommand{\keywords}[1]{\textbf{\textit{Keywords---}} #1}

\begin{document}

\title{Study of Anomaly Detection Based on Randomized Subspace Methods in IP Networks }
\name{\textnormal{Maboud F. Kaloorazi  ~and Rodrigo C. de Lamare } }


\address{{Centre for Telecommunications Studies (CETUC)}\\{Pontifical Catholic University of Rio de Janeiro, Brazil}\\
{Department of Electronics, University of York, United Kingdom}\\
Emails: {\{kaloorazi,~delamare\}}@cetuc.puc-rio.br
\vspace{-0.085em}}


\maketitle

\begin{sloppypar}

\begin{abstract}
In this paper we propose novel randomized subspace methods to detect
anomalies in Internet Protocol networks. Given a data matrix
containing information about network traffic, the proposed
approaches perform a normal-plus-anomalous matrix decomposition
aided by random subspace techniques and subsequently detect traffic
anomalies in the anomalous subspace using a statistical test.
Experimental results demonstrate improvement over the traditional
principal component analysis-based subspace methods in
terms of robustness to noise and detection rate. \\
\end{abstract}


\section{Introduction}
\label{Intro}

Network anomalies typically refer to abnormal behavior in the
network traffic such as traffic volume, bandwidth and protocol use,
which indicate a potential threat. Traffic anomalies may arise due
to various causes ranging from network attacks such as
denials-of-service (DoS) and network scans, to atypical
circumstances such as flash-crowds and failures, which can have
serious destructive effects on the performance and security of
Internet Protocol (IP) networks \cite{Thottan2003},
\cite{Zhang2005}.

The seminal paper by Lakhina et al. \cite{Lakhina2004} first
employed Principal Component Analysis (PCA) \cite{Jolliffe2002} to
detect network-wide traffic anomalies. Given a matrix of link
traffic data ${\bf Y}$, the approach performs a
normal-plus-anomalous matrix decomposition (i.e., ${\bf Y} =
\hat{\bf Y}+\tilde{\bf Y}$) using (a specific number of) its
principal components and seeks anomalies in the anomalous subspace
$\tilde{\bf Y}$. The emergence of this approach inspired researchers
to improve its performance and to evaluate its sensitivity for
detecting anomalies \cite{Ringberg2007}, \cite{Brauckhoff2009}.
Ringberg et al. \cite{Ringberg2007} point out that since PCA does
not consider the temporal correlation of the data, the normal
subspace is contaminated with anomalies. To address this issue,
Brauckhoff et al. \cite{Brauckhoff2009} propose to apply the
Karhunen-Loeve (KL) expansion \cite{Gray2005}, which considers both
the temporal and spatial correlations. Recently, inspired by the
well-established compressed sensing (CS) theory \cite{Donoho2006},
\cite{CRT2006} and also by robust principal component analysis
(RPCA) \cite{CSPW2009}, \cite{WPMGR2009}, \cite{CLMW2009}, several
works have approached network-wide traffic anomaly detection using
these methods (i.e., by solving a constrained optimization problem)
\cite{Roughan2012},
\cite{Mardani2016}. 

The PCA-based methods \cite{Lakhina2004}, \cite{Huang2007},
\cite{Brauckhoff2009} focus on link traffic covariance matrix and
accordingly compute its singular value decomposition (SVD), a
computationally expensive factorization, to separate the subspaces.
In this paper, we present two novel randomized subspace approaches
to detect anomalies in network traffic. In contrast to the works in
\cite{Lakhina2004}, \cite{Huang2007}, \cite{Brauckhoff2009}, the
proposed approaches do not form the covariance matrix and
consequently obviate the computation of the SVD for subspace
separation. We validate the proposed approaches using synthetically
generated data. Experimental results demonstrate that the proposed
techniques can successfully diagnose network-wide anomalies with
more effectiveness than PCA and robust PCA (RPCA).

The remainder of this paper is organized as follows. In
\refsec{Sigmodel} we introduce the signal model that represents IP
traffic and formulate the problem we are interested in solving. We
review the method of PCA for network anomaly detection in
\refsec{RPCAsec}. In \refsec{SRRPCAsec}, we describe our proposed
methods in detail. In \refsec{ERsec}, we present and discuss our
experimental results and our conclusion remarks are given in
\refsec{CONsec}.

\section{Signal Model and Problem Formulation}
\label{Sigmodel}

In this section, we describe a signal model that represents the
traffic in an IP network using linear algebra and state the problem
of interest. Based on the structure of a network and the flow of
data obtained by network tomography \cite{Vardi96}, we can model the
link traffic as a function of the origin-destination (OD) flow
traffic and the network-specific routing. Specifically, the
relationship between the link traffic ${\bf Y} \in \mathbb R^{m
\times t}$ and OD flow traffic ${\bf X} \in \mathbb R^{n \times t}$,
for a network with $m$ links and $n$ OD flows may be written as:
\begin{equation}
{\bf Y} = {\bf R}{\bf X},
\end{equation}
where $t$ is the number of snapshots and ${\bf R} \in \mathbb R^{m
\times n}$  is a routing matrix. The entries of ${\bf R}$, i.e.,
${\bf R}_{i,j}$, are assigned a value equal to one (${\bf
R}_{i,j}=1$) if the OD flow $j$ traverses link $i$, and are assigned
a value equal to zero otherwise.

The network traffic model that takes into account the anomalies and
the measurement noise over the links can be expressed by
\begin{equation}
\label{equation9} {\bf Y} = {\bf R}({\bf X} + {\bf A}) + {\bf V},
\end{equation} where ${\bf R} \in \mathbb R^{m \times n}$ is a fixed routing matrix,
${\bf X} \in \mathbb R^{n \times t}$ is the clean traffic matrix,
${\bf A} \in \mathbb R^{n \times t}$ is the matrix with traffic
anomalies and ${\bf V} \in \mathbb R^{m \times t}$ denotes the link
measurement noise samples. The problem we are interested in this
work is how detect anomalies by observing ${\bf Y}$.

\section{Principal Component Analysis for Network Anomaly Detection}
\label{RPCAsec}

Given the link traffic ${\bf Y}$, in order to detect anomalies the
work in \cite{Lakhina2004} performs a normal-plus-anomalous matrix
decomposition such that ${\bf Y} = \hat{\bf Y}+\tilde{\bf Y}$, where
$\hat{\bf Y}$ is the modeled traffic and $\tilde{\bf Y}$ is the
projection of ${\bf Y}$ onto the anomalous subspace
$\tilde{\mathcal{S}}$, using a selected number of its principal
components.

The modeled traffic represented by $\hat{\bf Y}$ is the projection
of ${\bf Y}$ onto the normal subspace $\mathcal{S}$ and the residual
traffic modeled by $\tilde{\bf Y}$ is the projection of ${\bf Y}$
onto the anomalous subspace $\tilde{\mathcal{S}}$. Specifically, the
modeled traffic can be obtained by
\begin{equation}
\label{equation2}
\hat{\bf Y} = {\bf P}{\bf P}^T{\bf Y}=\hat{\bf
C}{\bf Y}
\end{equation}
and
\begin{equation}
\label{equation2b}
 \tilde{\bf Y} = ({\bf I} - {\bf P}{\bf P}^T){\bf
Y}=\tilde{\bf C}{\bf Y},
\end{equation}
where ${\bf P} = [{\bf w}_1,{\bf w}_2,..., {\bf w}_r]$ is formed by
the first $r$ singular vectors of the covariance of the centered
traffic data $\hat{\bf \Sigma} = \frac{1}{t-1}({\bf{ Y-\mu}})({\bf
{Y-\mu}})^T$ and $\hat{\bf \Sigma} = {\bf W}{\bf \Lambda}{\bf W}^T$
is a singular value decomposition.

In order to detect abnormal changes in $\tilde{\bf Y}$, a statistic
referred to as the $Q$-statistic \cite{Jackson79} is applied by
computing the squared prediction error (SPE) of the residual
traffic:
\begin{equation}
{\text{SPE}} = {\|\tilde{\bf Y}\|_2^2} = {\|\tilde{\bf C}{\bf Y}\|_2^2},
\end{equation}
The network traffic is considered to be normal if
\begin{equation}
{\text{SPE}} \leq Q_\beta,
\end{equation}
where $Q_\beta$  is a threshold for the SPE defined as: 
\begin{equation}
Q_\beta = \theta_1\Big[\frac{c_\beta\sqrt{2\theta_2 h_0^2}}{\theta_1} +1 + \frac{\theta_2 h_0(h_0-1)}{\theta_1^2}\Big]^\frac{1}{h_0},
\end{equation}
where
\begin{equation}
\label{equation6} h_0 = 1-\frac{2\theta_1\theta_3}{3\theta_2^2}
\end{equation}
and
\begin{equation}
\theta_i = \sum_{j=k+1}^m \lambda_j^i, \text{for} \mspace{6mu}
i=1,2,3
\end{equation}
with $\lambda_j$ denoting the $j$-th singular value of $\hat{\bf
\Sigma}$ and  $c_\beta$ is the $1 - \beta$ percentile in a standard
normal distribution.

The singular vectors of $\hat{\bf \Sigma}$ (or principal components
of ${\bf Y}$) maximize the variance of the projected data. Thus, for
instance, the $j$-the singular value of $\hat{\bf \Sigma}$ (or the
variance captured by the  $j$-the PC) can be expressed as $\lambda_j
= \mathbb{V}{\text{ar}}\{({{\bf w}_j}^T {\bf Y})^T\}$. Note that, each
row in  ${\bf Y}$, $ Y_i \in \mathbb R^{1 \times t}$.

\section{Proposed Subspace-Projected Basis for Anomaly detection}
\label{SRRPCAsec}

This section describes our proposed approaches termed Randomized
Bases Anomaly Detection (RBAD) and Switched  Subspace-Projected
Bases for Anomaly Detection (SSPBAD). Similar to the works in
\cite{Dunia1998} and \cite{Lakhina2004}, given the data traffic
matrix ${\bf Y}$, RBAD and SSPBAD perform a normal-plus-anomalous
matrix decomposition. However, instead of the principal components
of ${\bf Y}$, they employ a matrix with a set of orthonormal bases
${\bf Q} \in \mathbb R^{m \times m}$ whose range approximates the
range of ${\bf Y}$. Once ${\bf Q}$ is constructed, as will be
explained in the next subsections, ${\bf Y}$ is represented as a
linear superposition of normal and anomalous components (${\bf Y} =
\hat{\bf Y}+\tilde{\bf Y}$) as given by
\begin{equation}
\hat{\bf Y} = {\bf P}{\bf P}^T{\bf Y}=\hat{\bf C}{\bf Y}
\end{equation}
and
\begin{equation}
\tilde{\bf Y} = ({\bf I} - {\bf P}{\bf P}^T){\bf Y}=\tilde{\bf
C}{\bf Y},
\end{equation}
where the matrix ${\bf P} = [{\bf q}_1,{\bf q}_2,..., {\bf q}_r]$
contains the first $r$ columns of ${\bf Q}$. Accordingly, the
variances captured by the orthonormal basis are computed as:
\begin{equation}
{\bf \Lambda_Q} = \mathbb{V}{\text{ar}}\{({{\bf Q}^T {\bf Y})^T\}}
\end{equation}
Then, the $Q$-statistic is applied to the anomalous component to
diagnose anomalies. In contrast to \cite{Dunia1998} and
\cite{Lakhina2004}, the proposed approaches do not require the
estimation of the covariance matrix from the data and, as a result,
the SVD is not required to be computed to separate subspaces. %
This also results in the reduction of the number of floating-point
operations (flops) to detect anomalies in the traffic network.

\subsection{Randomized Basis Anomaly Detection }

To separate normal and anomalous subspaces as in (\ref{equation2}),
RBAD uses orthonormal bases whose range approximates the range of
the traffic matrix ${\bf Y}$ (instead of the singular vectors of
$\hat{\bf \Sigma}$ used in \cite{Dunia1998} and \cite{Lakhina2004}).
To compute the bases, the product ${\bf B} = {\bf Y}{\bf \Phi}$ is
first formed using a random matrix ${\bf \Phi} \in \mathbb R^{t
\times m}$ and a $QR$ factorization is then performed on ${\bf B}$
(i.e., ${\bf Q}{\bf R}={\bf B}$) \cite{HMT2009}. To improve the
approximation accuracy the work in \cite{HMT2009} multiplies ${\bf
B}$ with ${\bf Y}$ and ${\bf Y}^T$ alternately. Once the bases are
obtained, the variances captured by ${\bf Q}$ are calculated (i.e.,
${\bf \Lambda_Q} = \mathbb{V}{\text{ar}}\{({{\bf Q}^T {\bf
Y})^T\}}$) to detect abnormal behavior in anomalous components.
Moreover, to apply $Q$-statistics the variances must be known
\cite{Jackson79}, \cite{Jackson91}. A pseudocode for RBAD is given
in Table \ref{TableOne}.

\begin{table}[!htb]
\caption{Pseudocode for the proposed RBAD technique.}\vspace{1em}
\label{TableOne}

\vspace{-1.6em}
\begin{center}
\rule{0.8\linewidth}{.2pt}
\end{center}
\vspace{-2em}
\noindent\hfil\rule{0.8\linewidth}{.2pt}\hfil

\begin{center}
\vspace{-0.5em}
\begin{minipage}{0.8\linewidth}

\renewcommand{\algorithmicrequire}{\textbf{Input:}}
\begin{algorithmic}[1]
\vspace*{-.3em}
\REQUIRE ~~ 
traffic matrix $\ {\bf Y} \in \mathbb R^{m \times t},
{\text{rank}}\mspace{6mu} r, {\text{an exponent}} \mspace{6mu} q $;
\vspace{1em}

        \STATE Generate a random matrix $\bf \Phi$;\vspace{1em}

        \STATE Form ${\bf B}=({\bf Y}{\bf Y}^T)^q {\bf Y \Phi}$;\vspace{1em}

        \STATE Perform a QR factorization to build an orthonormal basis: ${\bf B}={\bf Q}{\bf
        R}$;\vspace{1em}

        \STATE Compute the variances: \\ ${\bf \Lambda_Q} = \mathbb{V}{\text{ar}}\{({{\bf Q}^T {\bf
        Y})^T\}}$;\vspace{1em}

        \STATE Separate the subspaces with rank $r$: \\ ${\bf Y} = \hat{\bf Y}+\tilde{\bf
        Y}$;\vspace{1em}

        \STATE Apply $Q$-statistic to $\tilde{\bf Y}$: \\ if SPE $> Q_\beta \rightarrow$
        anomalies;\vspace{1em}

\RETURN anomalies in $\bf A$
\vspace*{-.3em}
\end{algorithmic}
\end{minipage}
\end{center}

\vspace{-1.3em}
\begin{center}
\rule{0.8\linewidth}{.5pt}
\end{center}
\vspace{-2em}
\noindent\hfil\rule{0.8\linewidth}{.5pt}\hfil

\end{table}

\subsection{Switched Subspace-Projected Basis for Anomaly Detection}

The proposed SSPBAD technique, similar to RBAD, also constructs
bases with orthonormal columns whose range approximates the range of
${\bf Y}$ which based on projects the traffic data ${\bf Y}$ onto
two subspaces orthogonal to each other ( $\hat{\mathcal{S}}$ and
$\tilde{\mathcal{S}}$). First, the product ${{\bf T}_1}= {{\bf
Y}^T}{{\bf T}_2}$ is formed using a random matrix ${{\bf T}_2} \in
\mathbb R^{m \times m}$. Next, ${{\bf T}_2}$ is updated by ${{\bf
T}_1}$ such that ${{\bf T}_2}= {\bf Y}{{\bf T}_1}$. Afterwards, a
$QR$ factorization is performed to construct the orthonormal bases
for the range of ${{\bf T}_2}$. These orthonormal bases will serve
as a surrogate to the bases of principal components used in
\cite{Dunia1998} and \cite{Lakhina2004} to separate normal and
anomalous subspaces. Subsequently, the variances captured by ${\bf
Q}$ are computed (i.e., ${\bf \Lambda_Q} =
\mathbb{V}{\text{ar}}\{({{\bf Q}^T {\bf Y})^T\}}$) to detect traffic
anomalies in the anomalous component using the $Q$-statistic.

A similar approach to constructing the orthonormal bases as in
SSPBAD was proposed in \cite{ZT2011} to approximate a rank-$r$
matrix, but they construct the bases for the range of ${{\bf T}_1}$.
To increase robustness of the algorithm for detecting anomalies, we
employ different matrices ${{\bf T}_2}$ as in \cite{KaDe16},
\cite{DeSa2009,mbsic,mbdf,aifir,sjidf,zhaocheng2012,mbdf2,armo,sbf,dce,dls,rwprec,arh,badstc}.
The random matrices generated include:
\begin{itemize}
  \item a matrix with i.i.d Gaussian entries i.e., $\mathcal{N}(0, 1)$,
  \item a matrix whose entries are i.i.d. random variables drawn from a Bernoulli distribution with probability 0.5,
  \item a Markov matrix whose entries are all nonnegative and the entries of each column add up to 1,
  \item a matrix whose entries are independently drawn from $\lbrace$-1, 1$\rbrace$.
\end{itemize}
Thus, SSPBAD switches among different random matrices and chooses
the best one in order to obtain the maximum number of anomalies. A
pseudocode for SSPBAD is given in Table \ref{TableTwo}.

\begin{table}[!htb]
\caption{Pseudocode for the proposed SSPBAD technique.}\vspace{1em}
\label{TableTwo} \vspace{-1.5em}
\begin{center}
\rule{0.85\linewidth}{.2pt}
\end{center}
\vspace{-2em}
\noindent\hfil\rule{0.85\linewidth}{.2pt}\hfil

\begin{center}
\vspace{-0.5em}
\begin{minipage}{0.8\linewidth}

\renewcommand{\algorithmicrequire}{\textbf{Input:}}
\begin{algorithmic}[1]
\vspace*{-.3em}
\REQUIRE ~~ 
 traffic matrix $\ {\bf Y} \in \mathbb R^{m \times t},
{\text{rank}}\mspace{6mu} r$;\vspace{1em}

\STATE Generate $N$ random matrices ${{\bf T}_2}$;\vspace{1em}
  \FOR{$i=$ 1: $N$}\vspace{1em}

  \STATE Form ${{\bf T}_1}$: ${{\bf T}_1}= {{\bf Y}^T}{{\bf T}_2}$;\vspace{1em}
  \STATE Update ${{\bf T}_2}$: ${{\bf T}_2}= {\bf Y}{{\bf T}_1}$;\vspace{1em}
 \STATE Perform a QR factorization to build orthonormal bases: ${\bf T}_2={\bf Q}{\bf R}$;\vspace{1em}
        \STATE Compute the variances: \\ ${\bf \Lambda_Q} = \mathbb{V}{\text{ar}}\{({{\bf Q}^T {\bf Y})^T\}}$;\vspace{1em}
        \STATE Separate the subspaces with rank $r$: \\ ${\bf Y} = \hat{\bf Y}+\tilde{\bf Y}$;\vspace{1em}
        \STATE Apply $Q$-statistic to $\tilde{\bf Y}$: \\ if SPE $> Q_\beta \rightarrow$ anomalies;\vspace{1em}

\ENDFOR \\\vspace{1em} \STATE Choose the best random matrix with
maximum number of anomalies;\vspace{1em} \RETURN anomalies in $\bf
A$ \vspace*{-.8em}
\end{algorithmic}

\end{minipage}
\end{center}

\begin{center}
\rule{0.8\linewidth}{.5pt}
\end{center}
\vspace{-2em}
\noindent\hfil\rule{0.85\linewidth}{.5pt}\hfil

\end{table}

\section{Experimental Results}
\label{ERsec}

To validate the proposed approaches, we conduct experiments on
synthetically generated data and compare them with PCA and RPCA. The
data matrix ${\bf Y}$ is generated according to the model in
(\ref{equation9}) with dimensions $m = 120, n = 240,t = 640$. The
low-rank matrix ${\bf X}$ is formed by a matrix multiplication ${\bf
U}{\bf V}^T$, where ${\bf U}\in \mathbb R^{n \times r}$ and ${\bf
V}\in \mathbb R^{t \times r}$  have Gaussian distributed entries
$\mathcal{N}(0, 1/n)$ and $\mathcal{N}(0, 1/t)$, respectively and $r
= 0.2\times m$. The routing matrix ${\bf R}$ is generated by entries
drawn from a Bernoulli distribution with probability $0.05$. The
sparse matrix of anomalies has $s=0.001\times mt $ non-zero elements
drawn randomly from the set $\{-1, 1\}$ and the noise matrix ${\bf
V}$ has independent and identically distributed (i.i.d) Gaussian
entries with variance $\sigma^2$, i.e., $\mathcal{N}(0, \sigma^2)$.
We set the confidence limit $1 - \beta=99.5\%$  for the value of the
$Q$-statistic for all three approaches.

\begin{figure}[ht]
\centering
\includegraphics[width=1\columnwidth]{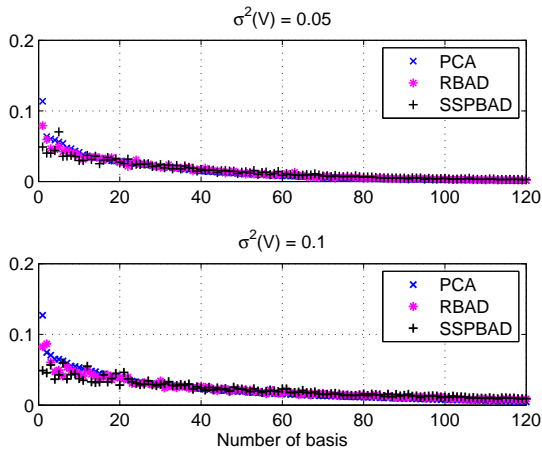}
\caption{ {A comparison of variances for PCA, RBAD, SSPBAD.}}
\label{figOne}
\end{figure}

In \reffig{figOne}, we compare the variances captured by the
proposed approaches (orthonormal basis) with the PCA method (PCs)
since they play a crucial role in the statistical test
($Q$-statistic) used to detect anomalies (cf. (\ref{equation6})). As
can be seen, returned variances by RBAD and SSPBAD are very close to
those returned by SVD.

\begin{figure}[ht]
\centering
\includegraphics[width=1\columnwidth]{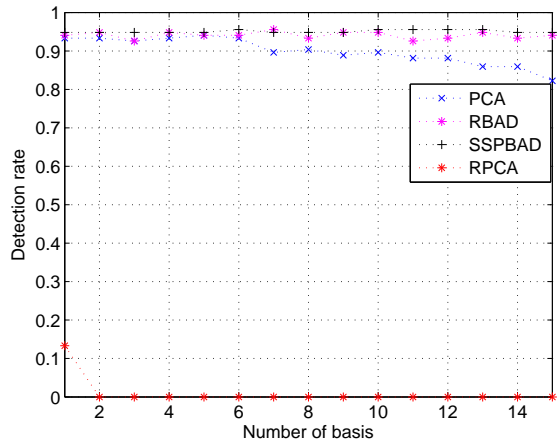}
\caption{ {A comparison of detection rate for PCA, RBAD, SSPBAD and
RPCA. Variance of the measurement noises $\sigma^2=0.1$}}
\label{figTwo}
\end{figure}

\reffig{figTwo} compares the detection rate against the number of
bases for different approaches.  As pointed out in \cite{Zhang2005}
the detection rate combines false-alarm rate and detection
probability into one measure and obviates the need for showing these
two probabilities in one versus the other manner. As can be seen,
the proposed RBAD and SSPBAD approaches outperform PCA when the
measurement noise has a higher variance. Furthermore, RPCA
\cite{CSPW2009},\cite{WPMGR2009}, \cite{CLMW2009} performs poorly.
Since we consider measurement noises ${\bf V}$ in our data model
(cf. \ref{equation9}), by increasing the rank, these noise samples
contaminate the matrix of outliers returned by RPCA and as a result
the abnormal patterns of the network (anomalies) cannot be
recovered.

\subsection{Computational Complexity}

The traditional PCA method operates on the link traffic covariance
($\hat{\bf \Sigma}$) to separate the subspaces. In particular, PCA
employs the SVD which requires $O(m^3)$ floating-point operations
(flops). RBAD and SSPBAD operate on the link traffic directly but
employ the $QR$ factorization, which requires $O(m^3)$ flops as
well. Although the computational complexity of RBAD and SSPBAD is
roughly the same as PCA in the context of anomaly detection, in
certain applications where SVD cannot be efficiently used, an
extension of the proposed approaches can be employed. For instance,
they can be used to build a direct solver for contour integral
equations with nonoscillatory kernels where the computational cost
for a $QR$ factorization is considerably less prohibitive than that
of SVD \cite{Cheng05}.

\section{Conclusion}
\label{CONsec}

In this paper, we have proposed the RBAD and SSPBAD random subspace
methods to detect traffic anomalies in IP networks. Both approaches
form normal and anomalous randomized subspaces by orthonormal bases
constructed for the range of the traffic data. A statistical test is
then applied and detects anomalies in the traffic. Simulations show
that RBAD and SSPBAD outperform PCA and RPCA. Future work will
concentrate on mathematical analysis of RBAD and SSPBAD.

{\small
\bibliographystyle{IEEEtran}
\bibliography{subssepref}}
\end{sloppypar}
\end{document}